\documentclass[thmsa]{article}
\usepackage{geometry}
\usepackage{graphicx}

\begin{document}

\author{G Arag\'{o}n-Gonz\'{a}lez, A. Canales-Palma, \and A. Le\'{o}%
n-Galicia, J. R. Morales-G\'{o}mez \\
PDPA. UAM- Azcapotzalco. Av. San Pablo \# 180. Col. Reynosa.\\
Azcapotzalco, 02800, D.F. Tel\'{e}fono y FAX: (55) 5318-9057.\\
e-mail: gag@correo.azc.uam.mx.}
\title{Maximum power, ecological function and efficiency of an irreversible
Carnot cycle. A cost and effectiveness optimization.}
\maketitle

\begin{abstract}
In this work we include, for the Carnot cycle, irreversibilities of linear
finite rate of heat transferences between the heat engine and its
reservoirs, heat leak between the reservoirs and internal dissipations of
the working fluid. A first optimization of the power output, the efficiency
and ecological function of an irreversible Carnot cycle, with respect to:
internal temperature ratio, time ratio for the heat exchange and the
allocation ratio of the heat exchangers; is performed. For the second and
third optimizations, the optimum values for the time ratio and internal
temperature ratio are substituted into the equation of power and, then, the
optimizations with respect to the cost and effectiveness ratio of the heat
exchangers are performed. Finally, a criterion of partial optimization for
the class of irreversible Carnot engines is herein presented.

\end{abstract}

\textbf{Nomenclature}

$I$: internal irreversible factor.

$Q$: heat transfer.

$q$: dimensionless heat transfer.

$W$: work.

$P$:\ power

$p$: dimensionless power.

$S$: entropy-generation rate.

$s$: dimensionless entropy-generation rate.

$K$: thermal conductance for heat loss.

$t$: time

$x$: internal temperatures ratio.

$y$: time ratio.

$z$: allocation ratio.

$U$: global heat transfer coefficient.

$A$: total heat transfer area.

$L$: thermal conductances ratio.

$C$: total cost.

Greek Symbols

$\alpha $: thermal conductance of hot side.

$\beta $: thermal conductance of cold side.

$\eta $ : efficiency.

$\sigma $: dimensionless dissipation.

$\epsilon $ : dimensionless ecological function.

$\mu $ : temperatures ratio of hot and cold sides.

Subscripts

$C$: Carnot.

$CI$: Carnot-like.

$CA$: endoreversible or Curzon-Ahlborn.

$H$: hot-side.

$L$: cold-side.

$max$: maximum.

$mp$: maximum power.

$me$: maximum efficiency.

$mec$: maximum ecological function.

Superscripts

$\ast $: second optimization.

$\ast \ast $: third optimization.

\section{Introduction}

The thermal efficiency of a reversible Carnot cycle is an upper limit of
efficiency for heat engines. In according to classical thermodynamics, the
Carnot efficiency is:%
\begin{equation}
\eta _{C}=1-\frac{T_{L}}{T_{H}}  \label{effC}
\end{equation}%
where $T_{L}$ and $T_{H}$ are the temperatures of the hot and cold
reservoirs between which the heat engine operates. The thermal efficiency $%
\eta _{C}$ can only be achieved through the infinitely slow process required
by thermodynamic equilibrium. Therefore, it is not possible to obtain a
certain amount of power output by using heat exchangers with finite heat
transfer areas. Thus, the thermal efficiency given in equation (\ref{effC})
does not have great significance and is a poor guide for the performances of
real heat engines.

A more realistic upper bound could be placed on the efficiency of a heat
engine operating at its maximum power point; the so-called CA efficiency
(Curzon-Alhborn \cite{curzon}):%
\[
\eta _{CA}=1-\sqrt{\frac{T_{L}}{T_{H}}} 
\]%
where the only source of irreversibility in the engine is a linear finite
rate heat transfer between the working fluid and its two heat reservoirs.

Real heat engines are complex devices. Besides the irreversibility of
finite-rate heat transfer in finite time taken into account in the
Curzon-Ahlborn engine (CA-engine), there are also other sources of
irreversibility, such as heat leaks, dissipative processes inside the
working fluid and so on. Thus, it is necessary to investigate more
comprehensively the influence of finite-rate heat transfer together with
other major irreversibilities on the performance of heat engines. For this
aim, we must consider general irreversible Carnot engines including three
major irreversibilities, which often exist in heat engines, and use it to
optimize the performance of an irreversible Carnot engine for several
objective functions.

In the past decade some new models of irreversible Carnot engines which
include other irreversibilities, besides thermal resistance, have been
established: heat leak and internal dissipations of the working fluid (see 
\cite{hoffman}, \cite{gordon}, \cite{chen}, \cite{yan} , \cite{ahmet}, \cite%
{arias}, \cite{calvo}, \cite{lchen}, \cite{aragon} and included references
there). Nevertheless, there are another parameters involved in the
performance and optimization of an irreversible Carnot cycle; for instance,
the allocation ratio of the heat exchangers, cost and effectiveness ratio of
the heat exchangers and so on (see \cite{calvo}, \cite{lchen} and \cite%
{lewins}).

In the optimization of Carnot cycles, including those irreversibilities, has
appeared four objective functions: power, efficiency, ecological and entropy
generation. The maximum power and efficiency have been obtained in \cite%
{chen}, \cite{yan} and \cite{aragon}. The maximum ecological function was
obtained in \cite{angulo} for the CA-engine and in form more general in \cite%
{arias}. Bejan \cite{bejan} has considered the minimization of the entropy
generation. In general, these optimizations were performed with respect to
only one characteristic parameter: internal temperature ratio.

In the first analysis of the CA-engine the time ratio of heat transfer from
hot to cold side was considered, but in further works this ratio was not
taken into account (see \cite{hoffman} for details). In this work this
relation is considered as a parameter. On the other hand, \cite{bejan} has
performed the optimization, also, with respect to other parameter: the
allocation ratio of the heat exchangers; and \cite{lewins} has considered as
parameters the cost and effectiveness ratio of the heat exchangers for the
CA-engine.

In this work we include, for the Carnot cycle, irreversibilities of linear
finite rate of heat transferences between the heat engine and its
reservoirs, heat leak between the reservoirs and internal dissipations of
the working fluid. A first optimization of the power output, the efficiency
and ecological function of an irreversible Carnot cycle, with respect to:
internal temperature ratio, time ratio for the heat exchange and the
allocation ratio of the heat exchangers; is performed. For the second and
third optimizations, the optimum values for the time ratio and internal
temperature ratio are substituted into the equation of power and, then, the
optimizations with respect to the cost and effectiveness ratio of the heat
exchangers are performed. Finally, a criterion of partial optimization for
the class of irreversible Carnot engines is herein presented.

This paper is organized as follows. In the section $2$ the relations for the
dimensionless power, efficiency, entropy generation and ecological function
of a class of irreversible Carnot engines are presented. In the section $3$,
the optimal analytical expressions for the efficiencies corresponding to
power and ecological function; and maximum efficiency are shown. In section $%
4$, the optimum values for the time ratio and internal temperature ratio are
substituted in the expression for dimensionless power. Then a second and
third optimizations of dimensionless power, are performed with respect to
the cost and effectiveness ratio of the heat exchangers. In the section of
Conclusions, a criterion of partial optimization for power, ecological
function, efficiency and entropy generation is presented.

\section{Irreversible Carnot engine.}

In considering the class of irreversible Carnot engines (see \cite{hoffman})
shown in Figure $1$, which satisfy the following five conditions:

\begin{figure}[t!]
\begin{center}
\includegraphics[width=10.0cm]{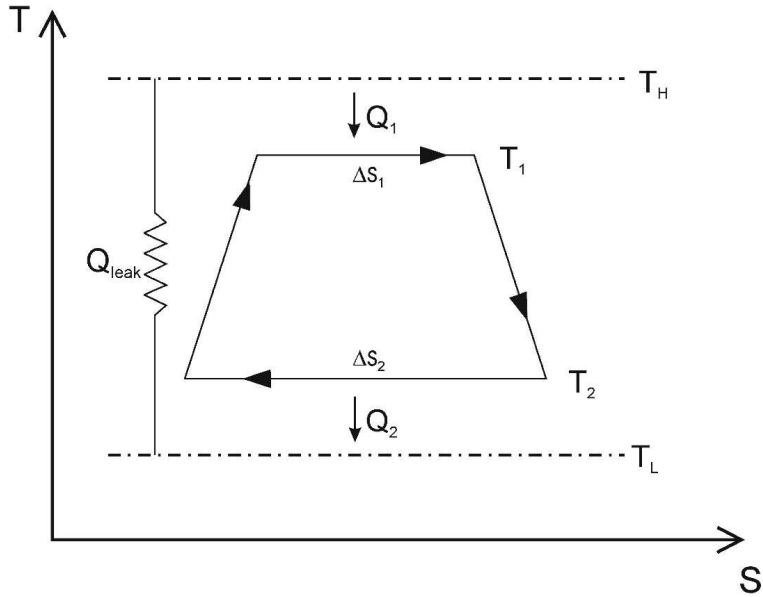}
\caption{A Carnot cycle with heat leak, finite-rate heat transfer and internal 
dissipations of the working fluid}
\end{center}
\end{figure}

(i) The cycle of the engine consists of two isothermal and two adiabatic
processes. The temperatures of the working fluid in the hot and cold
isothermal processes are, respectively, $T_{1}$ and $T_{2}$, and the times
of the two isothermal processes are, respectively, $t_{H}$ and $t_{L}$. The
temperatures of the hot and cold heat reservoirs are, respectively, $T_{H}$
and $T_{L}$.

(i) There is thermal resistance between the working fluid and the heat
reservoirs.

(ii) There is a heat lost $Q_{leak\textrm{ }}$from the hot reservoir to the
cold reservoir \cite{bejan}. In real engines heat leaks are unavoidable,
there are many features of an actual power plant which fall under that kind
of irreversibility, such as the heat lost through the walls of a boiler, a
combustion chamber, or a heat exchanger, and heat flow through the cylinder
walls of an internal combustion engine, and so on.

(iii) All heat transfer is assumed to be linear in temperature differences,
that is, Newtonian.

(iv) Besides thermal resistance and heat loss, there are other
irreversibilities in the cycle, the internal irreversibilities.\ For many
devices such as gas turbines, automotive engines, and thermoelectric
generator, there are other loss mechanisms, like friction or generators
losses, etc. that play an important role, but are hard to model in detail.
Some authors use the compressor (pump) and turbine isentropic efficiencies
to model the internal loss in the gas turbines or steam plants. Others, in
Carnot cycles, use simply one parameter to describe the internal losses.
Such a parameter is associated with the entropy produced inside the engine
during a cycle. Specifically, this parameter makes the Claussius inequality
becomes an equality (for details see \cite{hoffman}):%
\begin{equation}
\frac{Q_{2}}{T_{2}}-I\frac{Q_{1}}{T_{1}}=0  \label{claus}
\end{equation}%
where $I=\frac{\Delta S_{2}}{\Delta S_{1}}\geq 1$ (\cite{chen}).

Thus, the irreversible Carnot engine operates with fixed time $t$ allowed
for each cycle. The heat leakage $Q_{leak}$ is (\cite{bejan}): 
\[
Q_{leak}=K(T_{H}-T_{L})t 
\]

The heats $Q_{H},$ $Q_{L\textrm{ }}$transferred from the hot-cold reservoirs
are given by: 
\begin{eqnarray}
Q_{H} &=&Q_{1}+Q_{leak}=\alpha (T_{H}-T_{1})t_{H}+K(T_{H}-T_{L})t  \label{QH}
\\
Q_{L} &=&Q_{2}+Q_{leak}=\beta (T_{2}-T_{L})t_{L}+K(T_{H}-T_{L})t
\end{eqnarray}%
where $\alpha ,$ $\beta $ and $K$ are the thermal conductances and $%
t_{H},t_{L}$ are the time for the heat transfer in the isothermal branches,
respectively. The connecting adiabatic branches are often assumed to proceed
in negligible time (\cite{gordon}), such that the cycle contact total time $%
t $ is \cite{rubin}: 
\begin{equation}
t=t_{H}+t_{L}  \label{time}
\end{equation}

By first law and combining equations (\ref{QH}) and (\ref{claus}) we obtain :

\begin{equation}
W=Q_{1}(1-Ix)=\frac{T_{H}(1-Ix)\left( 1-\frac{\mu }{x}\right) }{\frac{1}{%
\alpha t_{H}}+\frac{I}{\beta t_{L}}}  \label{W}
\end{equation}%
\begin{equation}
Q_{H}=Q_{1}+Q_{leak}=\frac{W}{1-Ix}+K(T_{H}-T_{L})t  \label{QHf}
\end{equation}%
where $\mu =\frac{T_{L}}{T_{H}}$.\ And $x=\frac{T_{2}}{T_{1}}$ is a
characteristic parameter of the engine.

Now, the equation (\ref{time}) gives us the total time of the cycle, so it
can be parametrized as:%
\[
t_{H}=yt;\textrm{ }t_{L}=(1-y)t 
\]%
where $y=\frac{t_{H}}{t}=\frac{t_{H}}{t_{H}+t_{L}}$ is other characteristic
parameter of the engine.

Another parameter is the allocation of the exchangers heat \cite{bejan}. The
thermal conductances can be written as: 
\[
\alpha =UA_{H};\textrm{ }\beta =UA_{L} 
\]%
where $U$ is overall heat transfer coefficient and $A_{H}$ and $A_{L}$ are
the available areas for heat transfer. Then, an approach might be to suppose
that $U$ is fixed, the same for the hot side and the cold side heat
exchangers, and that the area $A$ can be allocated between both. The
optimization problem is then selected, besides of the optimum temperature
ratio and the time ratio, as the best allocation ratio. To take $UA$ as a
fixed value\ can be justified in terms of the area purchased, and the fixed
running costs and capital costs that altogether determine the overall heat
transfer coefficient (see (\cite{lewins})). Thus, for the optimization we
can take: 
\begin{equation}
\frac{\alpha }{U}+\frac{\beta }{U}=A  \label{rule1}
\end{equation}%
and parametrize it as:%
\[
\alpha =zUA;\textrm{ }\beta =(1-z)UA 
\]%
\begin{equation}
\frac{\alpha }{\beta }=\frac{z}{(1-z)}  \label{z}
\end{equation}

Therefore, the dimensionless power output, $p=\frac{W}{AUtT_{H}}$, and the
dimensionless heat transfer rate $q_{H}=$ $\frac{Q_{H}}{AUtT_{H}}$ are (by
equations(\ref{W}) and\ (\ref{QHf})):%
\begin{equation}
p=\frac{z\left( 1-z\right) y\left( 1-y\right) \left( 1-Ix\right) \left( 1-%
\frac{\mu }{x}\right) }{\left( 1-z\right) \left( 1-y\right) +zyI}
\label{powerIL}
\end{equation}%
\begin{equation}
q_{H}=\frac{z\left( 1-z\right) y\left( 1-y\right) \left( 1-\frac{\mu }{x}%
\right) }{\left( 1-z\right) \left( 1-y\right) +zyI}+L(1-\mu )  \label{qH}
\end{equation}%
where$\ L=\frac{K}{AU}$. And $z=\frac{\alpha }{UA}$ is the third
characteristic parameter of the engine. The thermal efficiency is given by:%
\begin{equation}
\eta =\frac{z\left( 1-z\right) y\left( 1-y\right) \left( 1-Ix\right) \left(
1-\frac{\mu }{x}\right) }{z\left( 1-z\right) y\left( 1-y\right) \left( 1-%
\frac{\mu }{x}\right) +L(1-\mu )\left( \left( 1-z\right) \left( 1-y\right)
+zyI\right) }  \label{eff_IL}
\end{equation}

The entropy-generation rate, $s_{gen}=\frac{S_{gen}}{AUtT_{H}}$, multiplied
by the temperature of the cold side, give us a dimensionless function $%
\sigma ,$ which is (equations (\ref{powerIL},\ref{qH})):%
\[
\sigma =T_{L}s_{gen}=T_{L}\left( \frac{q_{H}-p}{T_{L}}-\frac{q_{H}}{T_{H}}%
\right) =q_{H}(1-\mu )^{2}-p 
\]%
so,

\begin{equation}
\sigma =\frac{z\left( 1-z\right) y\left( 1-y\right) \left( 1-\frac{\mu }{x}%
\right) \left( Ix-\mu \right) }{\left( 1-z\right) \left( 1-y\right) +zyI}%
\allowbreak +L(1-\mu )^{2}  \label{entroIL}
\end{equation}

Finally, the ecological function \cite{angulo}, when $T_{L}$ is the
environmental temperature, is:

\[
\epsilon =p-\sigma =p\frac{2Ix-1-\mu }{Ix-1}+L(1-\mu ) 
\]%
$\allowbreak $then,%
\begin{equation}
\epsilon =\frac{z\left( 1-z\right) y\left( 1-y\right) \left( 1-\frac{\mu }{x}%
\right) \left( 1-2Ix\right) }{\left( 1-z\right) \left( 1-y\right) +zyI}%
+L(1-\mu )  \label{ecolIL}
\end{equation}%
when $I=1$ and $L=0$ the expressions for the CA-engine are obtained.

\section{Maximum power, ecological function and efficiency.}

In using the equation (\ref{powerIL}) and the extremes conditions: 
\[
\frac{\partial p}{\partial x}{\Huge |}_{(x_{mp},y_{mp},z_{mp})}=0;\textrm{ \ \ 
}\frac{\partial p}{\partial y}{\Huge |}_{(x_{mp},y_{mp},z_{mp})}=0;\textrm{\ }%
\frac{\partial p}{\partial z}{\Huge |}_{(x_{mp},y_{mp},z_{mp})}=0 
\]%
when the power reaches its maximum, $x_{mp},$ $y_{mp}$ and $z_{mp}$ are
given by:

\begin{equation}
x_{mp}=\sqrt{\frac{\mu }{I}}  \label{xmp}
\end{equation}%
\begin{equation}
y_{mp}=z_{mp}=\frac{1}{\sqrt[3]{I}+1}  \label{yzmp}
\end{equation}

Clearly $p$ reaches its maximum in $(x_{mp},y_{mp},z_{mp})$. Indeed, all the
critical points are (necessary condition): 
\[
\begin{tabular}{c}
$\left\{ z=0,y=y,x=\mu \right\} ,\left\{ x=\frac{1}{I},y=1,z=z\right\}
,\left\{ y=1,z=z,x=\mu \right\} $ \\ 
$\left\{ y=0,z=z,x=\mu \right\} ,\left\{ x=\frac{1}{I},z=0,y=y\right\}
,\left\{ y=0,z=0,x=x\right\} $ \\ 
$\left\{ z=1,y=1,x=x\right\} ,\left\{ z=1,y=y,x=\mu \right\} ,\left\{ x=%
\frac{1}{I},z=1,y=y\right\} $ \\ 
$\left\{ x=\frac{1}{I},y=0,z=z\right\} ,\left\{ x=\pm \sqrt{\frac{\mu }{I}}%
,y=\frac{1}{\sqrt[3]{I}+1},z=\frac{1}{\sqrt[3]{I}+1}\right\} $%
\end{tabular}%
\]

In eliminating the solutions without physical meaning, we see that there is
only one global critical point given by the equations (\ref{xmp}, \ref{yzmp}%
). Moreover, at this critical point maximum power developed. Indeed, a
sufficient condition for maximum power is, the eingenvalues of the Hessian ($%
\left[ \frac{\partial ^{2}p}{\partial w\partial u}|_{\left(
x_{mp},y_{mp},z_{mp}\right) }\right] _{w,u=x,y,z}$) must be negatives (\cite%
{panos}). It is clearly fulfilled that:

\[
\left[ 
\begin{array}{ccc}
\allowbreak -\frac{2\allowbreak I^{\frac{3}{2}}}{\sqrt{\mu }\left( 1+\sqrt[3]%
{I}\right) ^{3}} & 0 & 0 \\ 
0 & \allowbreak -\frac{2\left( 1-\sqrt{I\mu }\right) ^{2}}{\sqrt[3]{I}\left(
1+\sqrt[3]{I}\right) } & 0 \\ 
0 & 0 & -\frac{2\left( 1-\sqrt{I\mu }\right) ^{2}}{\sqrt[3]{I}\left( 1+\sqrt[%
3]{I}\right) }%
\end{array}%
\right] 
\]

The efficiency that maximizes the power $\eta _{mp}$ is given by (see
equation (\ref{eff_IL}) and Figure 2), 
\begin{equation}
\eta _{mp}=\frac{\left( 1-\sqrt{I\mu }\right) }{1+\frac{L\left( 1-\mu
\right) \left( \sqrt[3]{I}+1\right) ^{3}}{\left( 1-\sqrt{I\mu }\right) }}
\label{efmp}
\end{equation}

The generation of entropy is minimum when $y$ and $z$ are given by equation (%
\ref{yzmp}) and $x=\frac{\mu }{\sqrt{I}}.$ Nevertheless, for these values it
is seen that the corresponding power does not have physical meaning. For $x=%
\frac{\mu }{I}$ ($y=0,1$ or $z=0,1$)$,$ makes the first term of the equation(%
\ref{entroIL}) zero. The corresponding values of $y,z$ are also without
physical meaning. For $x=\mu $ ($y=0,1$ or $z=0$, $1)$ do not have physical
meaning either. Therefore, for this kind of Carnot engine, the entropy
generation does not have a global minimum within the valid interval. In \cite%
{lchen1} an engine that corresponds with the kind of irreversible Carnot
cycles herein presented is analyzed but the calculations leading to the
minimization of entropy generation are at fault, since they do not have
physical meaning. It results that the obtained power is negative! Thus, it
is only possible to minimize the entropy generation partially for the
variables $y,z$ and those values are given by: 
\begin{equation}
y_{m\sigma }=z_{m\sigma }=\frac{1}{\sqrt[3]{I}+1}  \label{yzment}
\end{equation}

In doing a analogous analysis for the ecological function, we have by the
equation (\ref{ecolIL}) that the unique critical point of ecological
function solutions with physical meaning is: 
\begin{equation}
x_{mec}=\sqrt{\frac{\mu \left( 1+\mu \right) }{2I}},  \label{xmec}
\end{equation}%
\begin{equation}
y_{mec}=z_{mec}=\frac{1}{\sqrt[3]{I}+1}  \label{yzmec}
\end{equation}%
and newly can see that its Hessian has all its negative eingenvalues.

The efficiency that maximizes the ecological function $\eta _{mec}$ is given
by (equation (\ref{eff_IL})):%
\begin{equation}
\eta _{mec}=\frac{\left( 1-\sqrt{\frac{\mu \left( 1+\mu \right) I}{2}}%
\right) }{1+\frac{L\left( 1-\mu \right) \left( \sqrt[3]{I}+1\right) ^{3}}{%
\left( 1-\sqrt{\frac{2I\mu }{\mu +1}}\right) }}  \label{efimec}
\end{equation}

Similarly, it's easily seen that there is only one critical point, with
physical meaning, for the efficiency, and it is given by: 
\begin{equation}
x_{me}=\frac{\mu +\sqrt{L\mu \left( 1-\mu \right) \left( 1+\sqrt[3]{I}%
\right) ^{3}\left( L\left( \sqrt[3]{I}+1\right) ^{3}\left( 1-\mu \right)
+1-I\mu \right) }}{I^{2}\left( L\left( \sqrt[3]{I}+1\right) ^{3}\left( 1-\mu
\right) +1\right) }  \label{xme}
\end{equation}

\begin{equation}
y_{me}=z_{me}=\frac{1}{\sqrt[3]{I}+1}  \label{yzme}
\end{equation}

To see, as above, that the efficiency reaches a maximum, becomes to
cumbersome a task if the solution of systems of equations are undertaken.
Therefore, an alternative way is presented in that follows, to obtain
equation (\ref{xme}). And when the efficiency reaches its maximum $\left(
x_{me},y_{me},z_{me}\right) $ is given by the equations(\ref{xme},\ref{yzme}%
).

Indeed, clearly the values of $y_{me},z_{me}$ given by the equation (\ref%
{yzme}) fulfill the following two extreme conditions:%
\[
\frac{\partial \eta }{\partial y}=0;\frac{\partial \eta }{\partial z}=0 
\]

Furthermore, as it was seen above, the optimal time ratio and the allocation
ratio are the same for both maximum power and ecological function (equations
(\ref{xmec}), (\ref{yzmec})). Therefore,%
\[
y_{mp}=y_{mec}=y_{me}=z_{mp}=z_{mec}=z_{me}=\frac{1}{\sqrt[3]{I}+1} 
\]

Thus, this values could be included in the equations of power and heat
transfer (equations (\ref{powerIL},\ref{qH})) and proceed to optimizes the
efficiency (equation (\ref{eff_IL}) by the following criterion valid when
there is only one parameter(\cite{aragon}):

\begin{description}
\item[Criterion (Maximum efficiency)] \textit{Let }$\eta =\frac{p}{q_{H}}$%
\textit{\ }$.$ \textit{Suppose \ }$\frac{\partial ^{2}p}{\partial x^{2}}%
|_{x}=\frac{\partial ^{2}q_{H}}{\partial x^{2}}|_{x},$\textit{\ for some }$%
x. $\textit{\ Then the maximum efficiency\ }$\eta _{\max }$\textit{\ is
given by } 
\begin{equation}
\eta _{\max }=\frac{\frac{\partial p}{\partial x}|_{x_{me}}}{\frac{\partial 
\mathit{\ }q_{H}}{\partial x}|_{x_{me}}}  \label{maxeff}
\end{equation}%
\textit{where }$x_{me}$\textit{\ is the point in which }$\eta $ \textit{%
achieves a maximum value.}
\end{description}

Then, by the equations (\ref{powerIL}) and (\ref{QH}) we obtain the
relationships of $p$ and $q_{H}$ with respect to $x.$%
\[
p=\frac{\left( 1-Ix\right) (1-\frac{\mu }{x})}{\left( \sqrt[3]{I}+1\right)
^{3}} 
\]%
\[
q_{H}=\frac{(1-\frac{\mu }{x})}{\left( \sqrt[3]{I}+1\right) ^{3}}+L(1-\mu ) 
\]

The conditions of the criterion are clearly satisfied. Indeed,%
\[
\frac{\partial ^{2}p}{\partial x^{2}}=\frac{\partial ^{2}q_{H}}{\partial
x^{2}}=\allowbreak -\frac{2\mu }{x^{3}\left( \sqrt[3]{I}+1\right) ^{3}}<0 
\]%
since $x>0.$Therefore (equation(\ref{maxeff})), 
\begin{equation}
\eta _{\max }=1-\frac{x_{me}^{2}I}{\mu }  \label{efirrmax}
\end{equation}%
where $x_{me}$ must, by the second law, satisfies the inequality (\cite%
{aragon}):

\begin{equation}
\textrm{\ \ }\frac{\mu }{I}\leq x_{me}\leq \sqrt{\frac{\mu }{I}}
\label{efimec1}
\end{equation}%
if we apply the preceding statement and the equation (\ref{efirrmax}), the
following inequality is obtained 
\begin{equation}
\eta _{mp}=1-\sqrt{I\mu }\leq \eta _{\max }\leq 1-I\mu =\eta _{CI}
\label{desefmaxcl}
\end{equation}%
where $\eta _{mp}=1-\sqrt{I\mu }$ and $\eta _{CI}=1-I\mu $ corresponding to
(Curzon-Ahlborn)-like and Carnot-like efficiencies; which includes the
internal irreversibilities in the $I$ factor.

Nevertheless, we can calculate easily $x_{me}$ from the following cubic
equation: 
\[
1-\frac{x_{me}^{2}I_{S}}{\mu }=\frac{p|_{x_{me}}}{q_{H}|_{x_{me}}}=\frac{%
(1-x_{me}I)(1-\frac{\mu }{x_{me}})}{(1-\frac{\mu }{x_{me}})+L(1-\mu )\left( 
\sqrt[3]{I}+1\right) ^{3}} 
\]%
In solving this equation and taking into account the inequality (\ref%
{efimec1}), we obtain the equation (\ref{xme}).

Finally, the maximum efficiency $\eta _{\max }$ is given by (equation (\ref%
{efirrmax}) ):%
\begin{equation}
\eta _{\max }=1-\left( \frac{\sqrt{I\mu }+\sqrt{L\left( 1-\mu \right) \left(
1+\sqrt[3]{I}\right) ^{3}\left( L\left( 1-\mu \right) \left( 1+\sqrt[3]{I}%
\right) ^{3}+1-I\mu \right) }}{1+L\left( 1-\mu \right) \left( \sqrt[3]{I}%
+1\right) ^{3}}\right) ^{2}  \label{effmay}
\end{equation}

The behavior of the efficiencies $\eta _{mp},\eta _{mec}$ and $\eta _{\max }$
is shown in the Figure 2.

In general it has been supposed that $I\geq 1;$ but sometimes can be
considered that $I=1$. In this case the internal irreversibilities can be
physically interpreted as part of the engine's heat leak that brings us to
the engine modeled in \cite{hoffman} and \cite{bejan}. So, substitution of $%
I=1$ into equations (\ref{xmp}; \ref{xmec}, \ref{xme}) and (\ref{yzme})
gives:%
\begin{eqnarray*}
x_{mp} &=&\sqrt{\mu };\textrm{ }x_{mec}=\frac{\left( 1-\sqrt{\frac{2\mu }{\mu
+1}}\right) \left( 1-\sqrt{\frac{\mu \left( 1+\mu \right) }{2}}\right) ^{2}}{%
8L\left( 1-\mu \right) +\left( 1-\sqrt{\frac{2\mu }{\mu +1}}\right) }\textrm{;}
\\
\textrm{ }x_{me} &=&\frac{\mu +\sqrt{8L\mu \left( 1-\mu \right) \left(
8L\left( 1-\mu \right) +1-I\mu \right) }}{8\left( 8L\left( 1-\mu \right)
+1\right) }
\end{eqnarray*}%
\[
y_{mp}=y_{mec}=y_{me}=z_{mp}=z_{mec}=z_{me}=\frac{1}{2} 
\]

\begin{figure}[t!]
\begin{center}
\includegraphics[width=10.0cm]{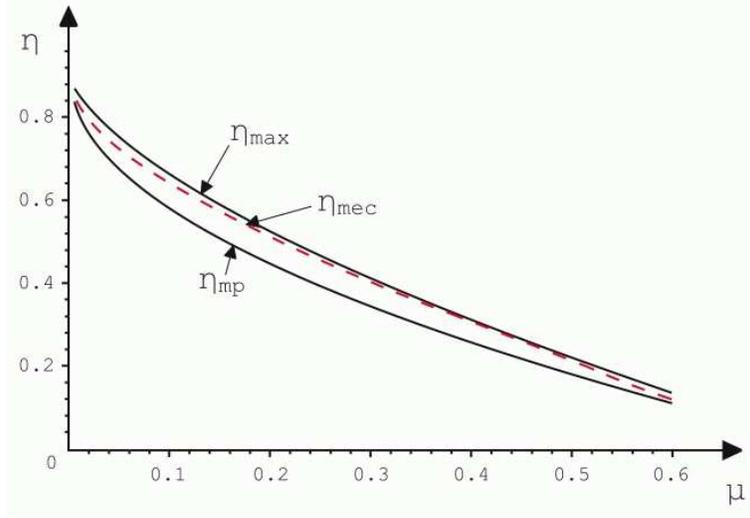}
\caption{Graphics of the efficiencies $\eta _{mp}$, $\eta _{mec}$ and 
$\eta _{\max }$ versus $\mu$ when $I=1.235$ and $L=0.01. $}
\end{center}
\end{figure}

The equations:%
\[
x_{mp}=\sqrt{\mu }\textrm{ and }z_{mp}=\frac{1}{2} 
\]%
are the same as the presented in \cite{bejan} and 
\[
x_{mp}=\sqrt{\mu }\textrm{ and }y_{mp}=\frac{1}{2} 
\]%
corresponding to the CA-engine. Further, the following results are obtained
(see equations (\ref{efmp}; \ref{efimec}, \ref{effmay})):%
\[
\eta _{mp}=\frac{1-\sqrt{\mu }}{2+\frac{8L\left( 1-\mu \right) }{1-\sqrt{\mu 
}}} 
\]%
\[
\eta _{mec}=\frac{1-\sqrt{\frac{\mu \left( 1+\mu \right) }{2}}}{1+\frac{%
8L\left( 1-\mu \right) }{\left( 1-\sqrt{\frac{2\mu }{\mu +1}}\right) }} 
\]%
\[
\eta _{\max }=1-\left( \frac{\sqrt{\mu }+\sqrt{8L\left( 1-\mu \right) \left(
8L\left( 1-\mu \right) +1-\mu \right) }}{1+8L\left( 1-\mu \right) }\right)
^{2} 
\]%
some of these results have been reported in the literature \cite{ahmet}.

On the other hand, the optimization performed in this work gives results
that could be applied to the design of power plants. For instance, in the
third section it is found that for 
\begin{equation}
y_{mp}=y_{m\sigma }=y_{mec}=y_{me}=z_{mp}=z_{m\sigma }=z_{mec}=z_{me}=\frac{1%
}{\sqrt[3]{I}+1}  \label{yzmax}
\end{equation}%
the engine operates at maximum power, efficiency and ecological function and
entropy generation local minimum. Therefore, the time rate in the isothermal
processes satisfies:%
\begin{equation}
\frac{t_{L}}{t_{H}}=\sqrt[3]{I}\geq 1  \label{ratet}
\end{equation}%
This result generalizes to one presented in \cite{hoffman}. And when $I=1,$%
\[
t_{L}=t_{H} 
\]

Similarly, it follows from the equation (\ref{yzmax}) that when the engine
operates at maximum power, efficiency and ecological function, the relation
for the heat transfer areas for the cold side to the hot side, is:%
\begin{equation}
\frac{A_{L}}{A_{H}}=\sqrt[3]{I}\geq 1;\frac{\beta }{\alpha }=\frac{UA_{L}}{%
UA_{H}}=\sqrt[3]{I}  \label{rateA}
\end{equation}%
This result shows that the size of the heat exchanger in the cold side must
be larger than the size of heat exchanger in the hot side. Thus, in
accordance with the definitions adopted for the thermal conductance, if $I>1$
the one for the cold side results greater than the hot side$.$ Furthermore,
if $I=1$%
\[
A_{L}=A_{H} 
\]%
which implies that the allocation of the heat exchangers is balanced (\cite%
{bejan}) .

By the equations (\ref{ratet}; \ref{rateA}) we have:%
\[
\frac{t_{L}}{t_{H}}=\frac{A_{L}}{A_{H}} 
\]%
which is satisfied when the heat engine operates to maximum power,
ecological function and efficiency, and minimum entropy generation. In \cite%
{aragon1} the above relationship was obtained, by a double optimization of
power and efficiency. For $I>1,$ the irreversibility produces an inverse
relationship between the total area and the total contact time; that is, a
less time is needed to transfer the heat that the engine processes. This is
due to the fact that less heat goes through the engine. Part of the heat is
lost because of internal irreversibility. For $I=1,$ the relationship
between area and contact time is inversely proportional; that is, if the
area is augmented the time is reduced. This result does not depend
explicitly of $I$ and differs of one presented in \cite{chen} and \cite%
{hoffman}.

\section{A cost and effectiveness optimization.}

A more detailed model would involve acknowledgement that the cost of
providing the same heat transfer capability differs between the cold and hot
sides. Let this represented as having a cost per unit heat transfer to be $%
c_{L}$ on the cold side but $c_{H}$ on the hot side (\cite{lewins}). Then%
\[
c_{H}\alpha +c_{L}\beta =C 
\]%
where $C$ is fixed total cost. Thus we have that the third characteristic
parameter $z_{1}$ changes to:%
\[
z^{\ast }=\frac{c_{H}\alpha }{C};1-z^{\ast }=\frac{c_{H}\beta }{C} 
\]%
In including the optimal values $x_{mp},y_{mp}$ (equations (\ref{xmp}, \ref%
{yzmp})) in the equation (\ref{W}) the dimensionless power $p^{\ast }=\frac{W%
}{C^{\ast }T_{H}t_{H}}$is given by: 
\[
p^{\ast }=\frac{\left( 1-\sqrt{I\mu }\right) ^{2}}{\frac{1}{z^{\ast }}+\frac{%
cI^{\frac{2}{3}}}{1-z^{\ast }}} 
\]%
where $\allowbreak c=\frac{c_{L}}{c_{H}}>1$ (equation(\ref{rateA})) and $%
C^{\ast }=\frac{C}{c_{H}}$. The efficiency is given by:%
\[
\eta ^{\ast }=\frac{\left( 1-\sqrt{I\mu }\right) ^{2}}{1-\sqrt{I\mu }%
+L\left( 1-\mu \right) \left( \sqrt[3]{I}+1\right) \left( \frac{1}{z^{\ast }}%
+\frac{cI^{\frac{2}{3}}}{\left( 1-z^{\ast }\right) }\right) } 
\]

In optimizing the power with respect to $z^{\ast },$ we have:%
\[
z_{mp}^{\ast }=\frac{1}{1+\sqrt{c}\sqrt[3]{I}} 
\]%
or equivalently%
\[
\frac{\beta }{\alpha }=\frac{\sqrt[3]{I}}{\sqrt{c}} 
\]%
Of course this reverts to the earlier form (equation(\ref{rateA})) if $c=1.$

The efficiency that develops maximum power is:%
\[
\eta _{mp}^{\ast }=\frac{1-\sqrt{I\mu }}{1+\frac{L\left( 1-\mu \right)
\left( 1+\sqrt[3]{I}\right) \left( 1+\sqrt[3]{I}\sqrt{c}\right) ^{2}}{1-%
\sqrt{I\mu }}} 
\]%
The Figure $3$ shows the behavior of $\eta _{mp}^{\ast }$versus $c$.

\begin{figure}[t!]
\begin{center}
\includegraphics[width=10.0cm]{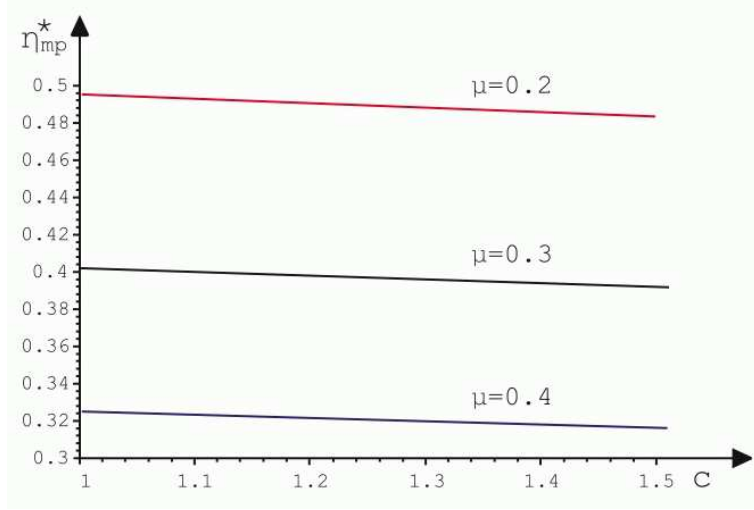}
\caption{Graphics of the efficiency $\eta _{mp}^{\ast }$ versus $c$
$(\mu =0.2,0.3,0.4)$ when $I=1.235$ and with $L=0.01.$}
\end{center}
\end{figure}

In general, there are two design rules for heat exchange at the two ends of
the heat engine (\cite{salah}). The first rule is that the thermal
conductance is constrained: 
\[
\alpha +\beta =\gamma 
\]%
where $\gamma $ is a constant; which was applied herein for the allocation
of the heat exchangers (see equation (\ref{z}) with $\gamma =UA$).

The second rule is that the total is constrained by 
\[
A=A_{H}+A_{L} 
\]%
where $A_{H},A_{L}$ are heat transfer areas on hot and cold side.

To apply the second rule, we may be faced with an existing heat exchange
apparatus which is to be redistributed between hot and cold sides to achieve
maximum power. Now, the total area $A$ is fixed but when distributed it has
different overall heat transfer coefficients and hence different
effectiveness on hot and cold sides. Thus, 
\[
A=A_{H}+A_{L}=\frac{\alpha }{U_{H}}+\frac{\beta }{U_{L}} 
\]%
where $U_{H},U_{L}$ are overall heat transfer coefficients on hot and cold
side. In parametrizing again.

\[
z^{\ast \ast }=\frac{\alpha }{U_{H}A};1-z^{\ast \ast }=\frac{\beta }{U_{L}A} 
\]%
Again, including the optimal values $x_{mp},y_{mp}$ (equations (\ref{xmp}, %
\ref{yzmp})) in the equation (\ref{W}) the dimensionless power $p^{\ast \ast
}=\frac{W}{AU_{H}t_{H}T_{H}}$is given by:%
\[
p^{\ast \ast }=\frac{\left( 1-\sqrt{I\mu }\right) ^{2}}{\frac{1}{z}+\frac{I^{%
\frac{2}{3}}}{\left( 1-z\right) U}} 
\]%
where $U=\frac{U_{L}}{U_{H}}$. And the efficiency is now given by:%
\[
\eta ^{\ast \ast }=\frac{\left( 1-\sqrt{I\mu }\right) ^{2}}{1-\sqrt{I\mu }%
+L\left( 1-\mu \right) \left( \sqrt[3]{I}+1\right) \left( \frac{1}{z}+\frac{%
I^{\frac{2}{3}}}{\left( 1-z\right) U}\right) } 
\]%
In optimizing the power with respect to $z^{\ast \ast },$ we have:%
\[
z_{mp}^{\ast \ast }=\frac{\sqrt{U}}{\sqrt[3]{I}+\sqrt{U}} 
\]%
or equivalently%
\[
\frac{\beta }{\alpha }=\sqrt[3]{I}\sqrt{U}\,;\textrm{ \ \ }\frac{A_{L}}{A_{H}}=%
\frac{\sqrt[3]{I}}{\sqrt{U}}=\sqrt[3]{I}\sqrt{\frac{U_{H}}{U_{L}}} 
\]%
Then, the optimal distribution of the heat exchangers areas is:%
\begin{eqnarray*}
A_{H} &=&\frac{A}{1+\sqrt[3]{I}\sqrt{\frac{U_{H}}{U_{L}}}}; \\
A_{L} &=&\frac{A}{1+\frac{1}{\sqrt[3]{I}}\sqrt{\frac{U_{L}}{U_{H}}}}
\end{eqnarray*}%
This result has been reported by \cite{kodal} (when $I=1)$ using another
thermoeconomic criterion. However, it differs when $I\neq 1$ \cite{kodal1}.

Then, the efficiency that develops maximum power is:%
\[
\eta _{mp}^{\ast \ast }=\frac{1-\sqrt{I\mu }}{1+\frac{L\left( 1-\mu \right)
\left( \sqrt[3]{I}+1\right) \left( 1+\allowbreak \frac{\sqrt[3]{I}}{\sqrt{U}}%
\right) ^{2}}{1-\sqrt{I\mu }}} 
\]

The Figure 4 shows the behavior of $\eta _{mp}^{\ast \ast }$ versus $U$.

\begin{figure}[t!]
\begin{center}
\includegraphics[width=10.0cm]{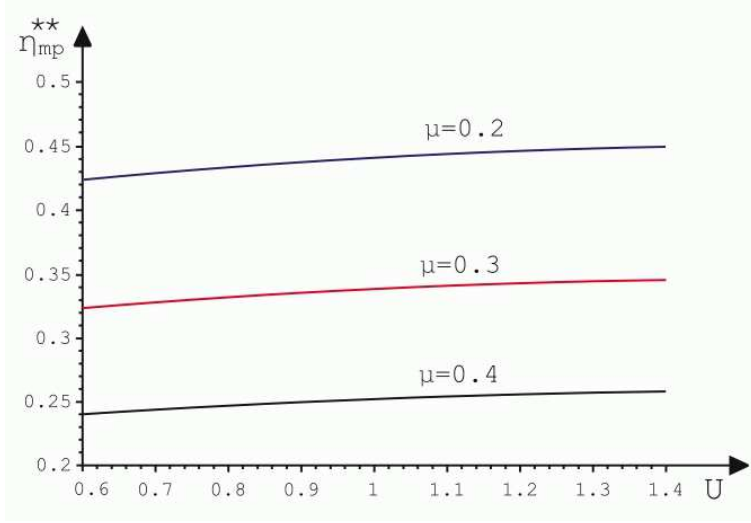}
\caption{Graphics of the efficiencies $\eta _{mp}^{\ast \ast }$
versus $U$ $(\mu =0.2,0.3,0.4)$ when $I=1.235$ and with $L=0.01$.}
\end{center}
\end{figure}

\section{Conclusion.}

In the above section, the optimization has been carried for one objective
function, that is, the power developed, respect to two additional
parameters. These parameters could be treated as economic ones. Moreover,
the obtained values, also optimizes the other considered objective
functions. Such as, entropy generation, ecological function and efficiency.
This is due to the fact that the analyzed Carnot engine satisfy the
following partial criterion:

\begin{description}
\item[Criterion] If $f_{i}=f_{i}(x,z,z^{\ast },z^{\ast \ast }...)$
represents one of the four objective functions, that is, power, efficiency,
ecological function or entropy generation; with (power, $i=1,...4;$ $x$ as
the internal temperature ratio; $z,z^{\ast },z^{\ast \ast }....$ are the
characteristic-economic parameters of Carnot cycles belonging to the class
of Carnot irreversible cycles analyzed. Moreover, if $z_{mj},z_{mj}^{\ast
},z_{mj}^{\ast \ast }...$are the optimum values for functions $f_{j}$ for
some $j$, then $\ z_{mj},z_{mj}^{\ast },z_{mj}^{\ast \ast }...$are also the
optimum values for the functions $f_{i},$ for $i\neq j.$
\end{description}

It is a fact that it suffices to develop the optimization for a couple of
objective functions, say the power and the efficiency to obtain the
parameters that optimizes the remaining objective functions.

For instance, the power and the efficiency satisfy the following functional
relationship (equation (\ref{eff_IL})):

\begin{equation}
\eta =\frac{p}{A+g(x)p}  \label{rel_func}
\end{equation}%
with $A=L(1-\mu )$ and $g(x)=\frac{1}{1-Ix}.$ Let $\phi =z,z^{\ast },z^{\ast
\ast }...$Then,%
\[
\frac{\partial \eta }{\partial \phi }=\frac{\frac{\partial p}{\partial \phi }%
\left( 1-\eta \right) }{A+g(x)p} 
\]%
Therefore,%
\[
\frac{\partial \eta }{\partial \phi }{\Huge |}_{\phi _{mp}=\phi
_{me}}=0\Longleftrightarrow \frac{\partial p}{\partial \phi }{\Huge |}_{\phi
_{mp}=\phi _{me}}=0 
\]

This implies that their roots are the same (necessary condition). It is
easily seen that for $\phi _{mp}=\phi _{me}$, the power and the efficiency
reach a maximum (sufficiency condition)

A remarkable conclusion of this work is that it is sufficient to find the
extreme of some of the functions $f_{i}$, say the power so that 
\[
\frac{\partial p}{\partial x}=0;\textrm{ }\frac{\partial p}{\partial \phi }=0 
\]%
where $\phi =z,z^{\ast },z^{\ast \ast }...$and then substitute in the
appropriate functional relationship (for the efficiency is equation(\ref%
{rel_func})) the values of $\phi _{mp}=\phi _{me}.$ The obtained $%
f_{i}=f_{i}\left( x\right) $ ($\eta =\eta (x)$) are then optimized respect
to the $x$ parameter only. It is found that the result optimizes the other
objective functions.

In other words, for the class of irreversible Carnot engines considered in
this work, the $x$ parameter could be considered as the fundamental
characteristic parameter of the engine. This is the only parameter that
changes its optimal value according to the engine operation conditions. The
remaining parameters maintain their optimal value independently of the
operation condition of the engine.

Finally, the above mentioned criteria could be applied and extended to other
models of irreversible engines \cite{ahmet}. Further work is underway.

\textbf{Acknowledgement: }This work was supported by the Program for the
Professional Development in Automation, through the grant from the
Universidad Aut\'{o}noma Metropolitana and Parker Haniffin - M\'{e}xico.

\end{document}